\documentclass{iau}
\usepackage{graphicx}
\usepackage[font={scriptsize}]{caption}

\def\kms{$\mathrm {km s}^{-1}$}
\newcommand{\arcmin}{$^{\prime}$}
\newcommand{\arcsec}{$^{\prime\prime}$}

\title[Balmer-dominated Shocks in Supernova Remnants] 
{An Integral View of Balmer-dominated Shocks in Supernova Remnants}

\author[Nikoli\'{c}, van de Ven, Heng, et al.] 
{Sladjana Nikoli\'{c}$^1$, Glenn van de Ven$^1$, Kevin Heng$^2$, Daniel Kupko$^3$,  
Jairo M\'{e}ndez-Abreu$^{4,5}$, J. Alfonso L. Aguerri$^{4,5}$, Joan Font Serra$^4$ \& John Beckman$^4$}

\affiliation{$^1$Max Planck Institute for Astronomy, K\"{o}nigstuhl 17, Heidelberg, Germany 
\\email: {\tt nikolic@mpia.de}\\[\affilskip]
$^2$University of Bern, Center for Space and Habitability, Sidlerstrasse 5, Bern, Switzerland \\[\affilskip]
$^3$Leibniz Institute for Astrophysics Potsdam (AIP), An der Sternwarte 16, Potsdam, Germany \\[\affilskip]
$^4$Instituto de Astrof\'{i}sica de Canarias (IAC), V\'{i}a L\'{a}ctea, La Laguna (Tenerife), Spain \\[\affilskip]
$^5$Departamento de Astrof\'{i}sica Universidad de La Laguna, E-38205 La Laguna (Tenerife), Spain
}

\pubyear{2013}
\volume{296}  
\pagerange{--}
\jname{Supernovae environmental impact}
\editors{Alak Ray \& Dick McCray}
\begin{document}

\maketitle

\begin{abstract}
We present integral-field spectroscopic observations with the
VIMOS-IFU at the VLT of fast (2000-3000\,\kms) Balmer-dominated
shocks surrounding the northwestern rim of the remnant of
supernova 1006. The high spatial and spectral resolution of the instrument
enable us to show that the physical characteristics of the shocks exhibit
a strong spatial variation over few atomic scale lengths across 133 sky locations. 
Our results point to the presence of a population of non-thermal protons 
(10-100\,keV) which might well be the seed particles for generating high-energy cosmic rays. 
We also present observations of Tycho's supernova remnant taken with the narrow-band
tunable filter imager OSIRIS at the GTC and the Fabry-Perot interferometer
GHaFaS at the WHT to resolve respectively the broad and narrow H$\alpha$ lines across 
a large part of the remnant.
\end{abstract}

\firstsection 
\section{Introduction}

Supernova remnants are laboratories for studying optical shocks.
An optical spectrum dominated by Balmer lines is seen when a fast
astrophysical shock enters partly neutral interstellar gas. Balmer
dominated shocks (BDSs) are characterized by velocities higher
than 200\,\kms , the presence of two-component H$\alpha$ line, the
absence of forbidden lines of metals in low ionization states, and general
lack of non-thermal X-ray emission at the location where H$\alpha$
lines are detected (\cite[Heng 2010]{Heng10}). BDSs are observed around
historical supernova remnants (SNRs)
like Tycho, Kepler and SN1006. \\
\indent The hydrogen lines consist of a narrow ($\sim$10\,\kms) and
a broad ($\sim$1000\,\kms) component. The narrow component is
produced by cold hydrogen atoms in the pre-shock ambient interstellar
medium (ISM) that are collisionally excited by electrons and
protons in the shock. While the broad component is produced by
post-shock hot neutrals, created through charge exchange between
incoming hydrogen atoms and hot protons in the shock. Thus, the
two-component H$\alpha$ lines directly yield the pre-shock and
post-shock temperatures of the ISM around the remnant. The Balmer
line profiles also contain signatures of shock precursors.
Investigating in detail the shape of the H$\alpha$ line has the
potential to provide strong observational constraints on cosmic
rays (CRs). In particular, the CRs will heat the cold hydrogen atoms in
the ISM before they are being ionized by the shock, resulting in a
narrow H$\alpha$ line of which the width is broadened beyond the
normal 10-20\,\kms gas dispersion. The CRs can also carry away
energy from the protons in the post-shock, so that the broad
H$\alpha$ line has a smaller width than allowed by the fast shock
velocity. Because the CR precursor is typically spatially
unresolved, its additional contribution to the narrow-line
H$\alpha$ emission results in a decreasing broad-to-narrow
intensity line ratio. The presence of CRs can also cause the shape
of the broad H$\alpha$ line to deviate from a Gaussian profile
(\cite[Raymond \etal\ 2010]{Raymond10}).\\
\indent Here we present high-spatial resolution spectro-photometric imaging
of the remnants SN\,1006 and Tycho, and investigate the H$\alpha$-line profiles in detail.

\section{Data \& Analysis}

\subsection{VIMOS-IFU Observations of SN\,1006}

\indent
The long-slits typically used to measure the H$\alpha$ line have
low spatial resolutions resulting in the contribution of multiple shock
fronts to the measured H$\alpha$ line. It is then unclear
whether the line shape, including its width, are contaminated by this
geometric effect. Optical integral-field unit (IFU) spectrographs
are able to trace and distinguish multiple, projected shocks.
Since BDSs have not been investigated before with IFUs, our aim in
this pilot study was to demonstrate that such observations can be
executed and that the scientific yields constitute a marked
improvement over previous studies.

\begin{figure}[h!]
\centering
\includegraphics[width=0.4\textwidth,angle=-90]{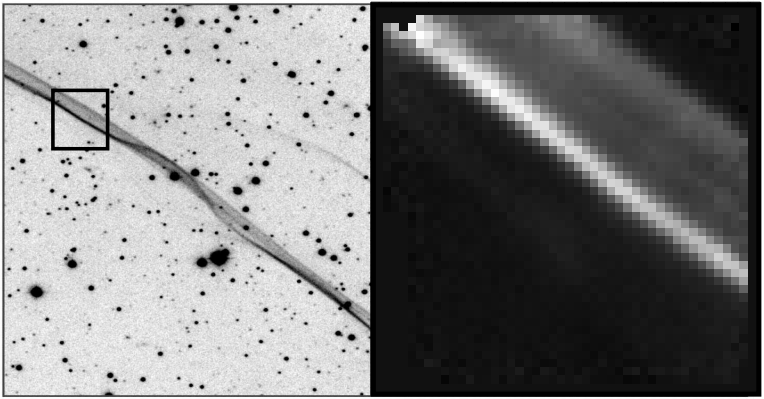}
\caption{VIMOS-IFU spectroscopy of the shock front in the remnant of SN1006. The left panel shows
a CTIO-Curtis-Schmidt narrow-band H$\alpha$ image (from Winkler \etal\ 2003) of the NW rim of the
remnant with the box indicating the region observed with the VIMOS-IFU. The right panel
shows the reduced data cube collapsed in wavelength around the narrow H$\alpha$ line, nicely recovering 
the shock in the image.}
\label{fig:sn1006_vimos}
\end{figure}
\begin{figure}[h!]
\hspace{0.2in}
\includegraphics[width=0.5\textwidth,angle=-90]{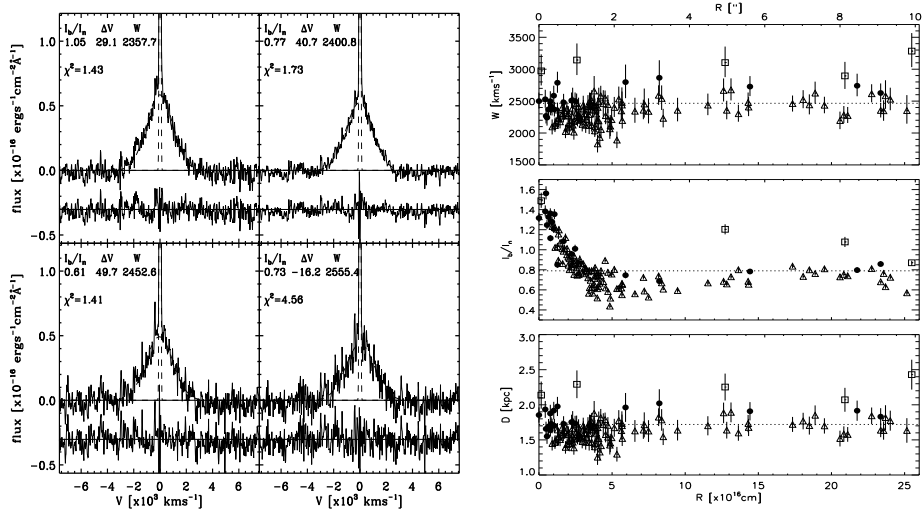}
\vspace{-0.1in}
\caption{Two-component H$\alpha$-line profiles and spatial variations of shock parameters.
The left four panels show the spectra that have been combined from four different regions
parallel to the shock front. In each panel, we show the best-fit parameters of the double-Gaussian fitting 
(dashed lines): broad-to-narrow intensity ratio
$I$, velocity offset $\Delta$V from the narrow component (in \kms), and the FWHM of the broad component(in \kms).
The three panels on the right show 133 measured (with error bars) broad line widths $W$, broad-to-narrow intensity ratios
$I_{b}/I_{n}$, and heliocentric distances $D$ from inferred shock velocities combined with measured proper motions 
of the shock front (\cite[Winkler \etal\ 2003]{Winkler03}). Data are ordered in increasing distance from the inner 
rim to the outer rim. The dashed horizontal lines indicate the measured $W$=2465.76 \kms\ and $I_{b}/I_{n}$=0.79 
values from collapsing all spectra of the pixels on the shock front, and from there the inferred $D$=1.72\,kpc. 
Data points marked with open triangles (squares) indicate that the $I_b/I_n$ values are too low (high) to obtain 
a solution of the van Adelsberg \etal\ model that does not include non-thermal physics.}
\vspace{-0.1in}
\label{fig:sn1006_spatial}
\end{figure}

\vspace{-0.1in}
\indent We used the VIMOS-IFU spectrograph on the Very Large
Telescope (VLT) to investigate the fast optical shocks at the
northwestern (NW) rim of the remnant of SN1006 (see Figure~\ref{fig:sn1006_vimos}).
The 27\arcsec$\times$27\arcsec\
field-of-view (FOV) of the VIMOS-IFU on a 8\,m telescope enables us
to collect enough photons in a reasonable time to reach the high
signal-to-noise ratio (S/N) required to accurately measure the
shape of the H$\alpha$ line, including deviations from a Gaussian
profile. The high spatial resolution of 0.\arcsec67
($\simeq$2$\times$10$^{16}$\,cm at the distance of $\simeq$2\,kpc), in combination
with the two-dimensional coverage of the VIMOS-IFU allow us to
precisely trace the narrow shock front. The spectral resolution
with dispersion of $\backsim$48\,\kms\ is more than sufficient to
measure the width of the broad line of 2000-3000\,\kms, as well as
deviations from a Gaussian profile. The width of the narrow line
of $\backsim$20\,\kms can not be resolved, but its intensity can
be accurately measured.\\
\indent
As a first analysis we combined
the spectra from four different regions parallel to the shock front (for coloured figures see Nikolic \etal\ 2013) 
and analyzed the resulting two-component H$\alpha$ lines shown in Figure~\ref{fig:sn1006_spatial}.
Performing double-Gaussian fits we extracted various parameters, including the
broad-to-narrow line ratio ($I_{b}/I_{n}$), velocity offset ($\Delta$V) from the
narrow component and the FHWM of the broad-line component ($W$).
Already strong variation in $I_{b}/I_{n}$ and $W$, as well as hint of non-Gaussianity 
in the broad-line core point towards the presence of a precursor in the shock.

\indent In order to investigate spatial variations of the physical characteristics across the shock front,
we used the method of Cappellari \& Copin (2003) to combine the neighboring pixels
and create 133 spatial (Voronoi) bins in which the combined spectra have a minimal S/N of 40.
We fitted double-Gaussian to the lines and extracted again the same parameters (per bin)
as before: $I_{b}/I_{n}$, $\Delta$V and $W$. We then converted for
each bin $W$ and $I_{b}/I_{n}$ to a shock velocity and electron-to-proton temperature
ratio ($\beta$) using the model of van Adelsberg \etal\ (2008). The variations of the intensity ratios
and broad-line widths along with the distances calculated from derived shock velocities and the
proper motion measurements of 280\,masyr$^{-1}$ (\cite[Winkler \etal\ 2003]{Winkler03})
are shown in the three right panels of Figure~\ref{fig:sn1006_spatial}.
Nearly 85\% of the observed values (represented with empty triangles and squares) are out of 
the range predicted by the model, which does not include CR physics.
Strong spatial variations in the intensity ratios and broad-line widths, the fact that the most
of our observed values are not in the model predicted range, and the hint
of the non-Gaussianity in the broad-line core all together indicate the potential presence of suprathermal protons,
non-thermal particles which might be the seed particles for generating high-energy CRs (\cite[Nikolic \etal\ 2013]{Nikolic13}).

\subsection{OSIRIS \& GHaFaS Observations of Tycho}

\indent
Integral-field spectrographs are able to trace and distinguish multiple, projected shocks, but still,
the field-of-view of these spectrographs typically covers only a small portion of the full remnant.
Henceforth, we use OSIRIS narrow-band tunable filter imaging at GTC (Gran Telescopio Canarias). 
The high spatial resolution of 0.\arcsec125 per pixel avoids geometric effects, while the large 
field-of-view of 4\arcmin$\times$4\arcmin\ allows to measure the width of the broad component of the 
H$\alpha$ line of the shock fronts in the full northeastern part of the remnant of Tycho's supernova.
Our goal is to map the changing CR acceleration efficiency along the remnant by measuring broad H$\alpha$-line
widths smaller than expected from the high shock velocities, which we know independently from proper motion 
measurements of the shock fronts.\\
\indent
OSIRIS low spectral resolution prevents resolving the narrow H$\alpha$ component. In order to precisely measure
the intrinsic narrow H$\alpha$-line width, we use the Fabry-Perot interferometer GHaFaS at
the WHT (William Herschel Telescope) which has the unique capability to scan with a spectral resolution
as fine as 8\,\kms. This enables us to directly quantify the presence of the CR precursor.
The high spatial resolution of 0.\arcsec2 plus large FOV 3.\arcmin4$\times$3.\arcmin4 of GHaFaS allow us to 
differentiate between different individual shock fronts along Tycho's supernova remnant. Observing exactly the 
same parts of the Tycho's remnant with both OSIRIS and GHaFaS, we expect to place tight constrains on CR precursors 
in Tycho's SNR.

\begin{figure}[ht]
\includegraphics[width=0.5\textwidth,angle=-90]{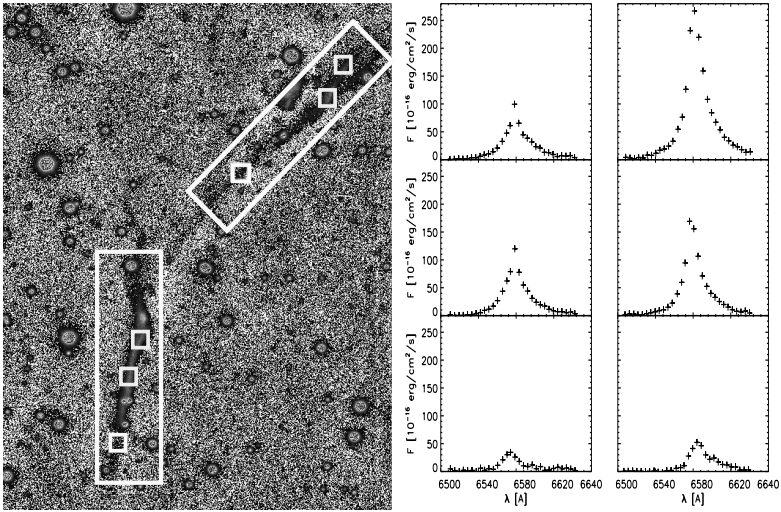}
\centering
\caption{OSIRIS observations of Tycho's remnant. The left panel shows reduced OSIRIS narrow-band imaging data of the northeastern rim
of the remnant. Small 5$\times$5 pixel boxes (white squares) indicate six different positions along the nicely recovered shock front
from which the H$\alpha$ line was extracted and shown in the right panel. The first column shows the spectra from the upper white
rectangular region from the top to the bottom, while the second column shows the spectra from the lower white rectangular region again
in the same order.}
\vspace{-0.2in}
\label{fig:tycho}
\end{figure}

\vspace{0.1in}
\noindent
\textbf{Acknowledgments}\\
\indent
We would like to thank Bernd Husemann (AIP), John C. Raymond (Harvard-Smithsonian CfA), John P. Hughes (Rutgers University) 
\& Jes\'{u}s Falc\'{o}n-Barroso (IAC) for their collaboration in the work of VIMOS-IFU observations of SN\,1006.

\end{document}